\documentclass[traditabstract]{aa} 
\usepackage{longtable}
\usepackage{amsmath}
\usepackage{graphicx}
\usepackage{txfonts}
\usepackage{natbib}
\usepackage{verbatim}
\begin{document}
   \title{Spectroscopic stellar parameters for 582 FGK stars in the HARPS volume-limited sample}
  \subtitle{Revising the metallicity-planet correlation}

   \author{S. G. Sousa\inst{1,}\inst{2}
          \and
	  N. C. Santos\inst{1,}\inst{3,}\inst{4}
	  \and G. Israelian\inst{2,}\inst{5}
	  \and M. Mayor\inst{3}
	  \and
	  S. Udry\inst{3} 
          }

	  \institute{Centro de Astrof\'isica, Universidade do Porto, Rua das Estrelas, 4150-762 Porto, Portugal
	  \and Instituto de Astrof\'isica de Canarias, 38200 La Laguna, Tenerife, Spain
	  \and Geneva Observatory, Geneva University, 51 Ch. des Maillettes, 1290 Sauverny, Switzerland
	  \and Departamento de F\'isica e Astronomia, Faculdade de Ci\^encias da Universidade do Porto, Portugal
	  \and Departamento de Astrofisica, Universidade de La Laguna, E-38205 La Laguna, Tenerife, Spain
}

   \date{}

 
\abstract{
To understand the formation and evolution of solar-type stars and planets in the solar neighborhood, we need to obtain their stellar parameters with high precision.
We present a catalog of precise stellar parameters for low-activity FGK single stars in a volume-limited sample followed by the HARPS spectrograph in the quest to identify extra-solar planets. 
The spectroscopic analysis was completed assuming LTE with a grid of Kurucz atmosphere models and using the ARES code to perform an automatic measurement of the line equivalent widths.
The results are compared with different independent methods and also with other values found in the literature for common stars. Both comparisons are consistent and illustrate 
the homogeneity of the parameters derived by our team.
The derived metallicities of this sample reveal a somewhat different distribution for the present planet hosts, but still indicates the already known higher frequency of planets 
observed for the more metal-rich stars. We combine the results derived in this sample with the one from the CORALIE survey to present the largest homogeneous spectroscopic study
of the metallicity-giant-planet relation using a total of 1830 stars.

}

\keywords{Stars: fundamental parameters – planetary systems – Stars: abundances – Stars: statistics}
\authorrunning{Sousa, S. G. et al.}
\titlerunning{Spectroscopic characterization of a volume limited sample...}

\maketitle

\section{Introduction}

More than half of a thousand planets have been detected since the discovery of the first exoplanet orbiting a solar-type star \citep[][]{Mayor-1995}. The method that has been adopted the most widely in achieving these  
detections is the radial velocity technique. There are several dedicated observing programs that are, almost, continuously monitor the sky. The data obtained by these programs 
consists of stellar spectra collected by different high resolution spectrographs located at many different observatories around the world.
Today the HARPS spectrograph \citep[][]{Mayor-2003} is at the top of the ``food-chain'' when talking about planet hunting. This is not only due to its high spectral resolution but more importantly its long-term stability in the determination 
of the radial velocity of stars.

These new discoveries of planets are allowing us to constrain the theory of formation and evolution of planetary systems. Soon after the first detections of planets, it was found that the stars hosting 
the newly discovered giant planets were systematically more metallic than a normal sample of stars. This well-established correlation has been confirmed by 
several authors \citep[][]{Gonzalez-1997, Gonzalez-2001, Santos-2001, Santos-2004b, Fischer_Valenti-2005, Udry-2006, Udry-2007b}, and has been shown to imply that core accretion is the main mechanism for the formation of 
giant planets \citep[][]{IdaLin-2004,Mordasini-2009} and not disk-instability\citep[][]{Boss-2002}. 
This observational correlation might also be consistent with more recently developed theories such as the formation of planets by the tidal downsizing of giant planet embryos \citep[][]{Nayakshin-2010}.

In this paper, we present a catalog of spectroscopic stellar parameters for FGK single stars of low activity that belong to the volume-limited sample observed with HARPS to search planets. 
In Sect. 2, we describe briefly the HARPS GTO sample analyzed in this work. Section 3 describes the procedure used to derive precise spectroscopic stellar parameters 
as well as estimate the stellar masses. In Sect. 4, we compare our derived parameters with other independent methods described by our team in previous works. 
In Sect. 5, we compare our spectroscopic derived parameters with others that we found in the literature to check consistency between different methods, while In Sect. 6, we present the solar-twins that 
are present in the analyzed sample. In Sect. 7, we combine this HARPS sample with the initial CORALIE sample. We compare both to check for differences in the [Fe/H] distributions. These two samples combined form 
the largest volume-limited sample of single stars of low activity with homogeneous and precise measurements of [Fe/H] that we use to study the metallicity correlation in planet host stars. Finally, in Sect. 8 
we summarize the work presented here.

\section{The sample \& observations}

The sample of stars presented here is part of a HARPS GTO program that aims to detect and obtain accurate orbital elements of Jupiter-mass planets in a well-defined volume of the solar neighborhood. 
Although giant planets were the main targets of this program, there is also the possibility of detecting Neptune-mass planets. This sample was compiled with stars in the solar neighborhood out to 
57.5 pc \citep[][]{LoCurto-2010} from the Sun 
to complete a planet-search survey previously started with the CORALIE spectrograph \citep[][]{Udry-2000}. The selection of the stars within this limited volume (containing more than 1400 stars) was carried out to ensure that the stars are suitable 
to obtain precise radial velocities. They have a wide range of spectral types extending from the cooler M0 to the hotter F2, although the majority are solar-type stars. All the stars were selected to have low levels of activity and low rotation 
rates. The sample was also clean of known binaries and variable stars. The total number of stars within this program reaches $\sim$ 850.

This sample indeed represents an extension of the one studied by the CORALIE program, and collectively they form the largest volume-limited sample with homogeneous measurements of [Fe/H]. In the past few years there have been 
some discoveries of 
planets orbiting the stars in this sample, including also a few that were posteriorly unconfirmed \citep[][]{Tinney-2003, Hebrard-2010, Santos-2010b, Naef-2010, Moutou-2009, LoCurto-2010, Santos-2011, Mordasini-2011, Moutou-2011}. 
These will allow us to increase the number of stars with planets in a subsample to ensure a proper statistical comparison of the metallicity distributions.

\begin{figure}[!t]
\centering
\includegraphics[width=8cm]{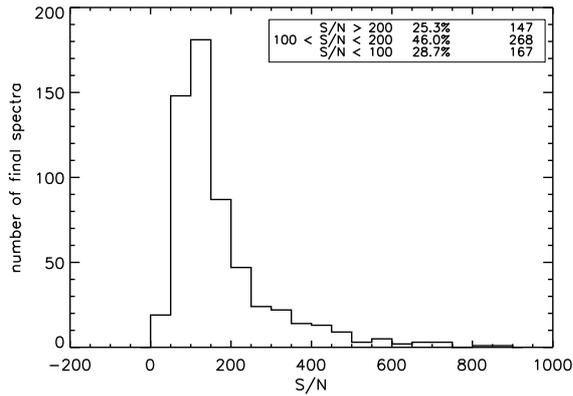}
\caption[]{Distribution of the signal-to-noise ratio of the final spectra compilation.}
\label{fig1}
\end{figure}

The spectra for the sample were collected by HARPS during several runs, thus there was the need to combine the spectra of each star into one to achieve a higher signal-to-noise (hereafter S/N). 
The individual spectra of each star were reduced 
using the HARPS pipeline and combined using IRAF \footnote{IRAF is distributed by National Optical Astronomy Observatories, operated by the Association of Universities for Research in 
Astronomy, Inc., under contract with the National Science Foundation, U.S.A.} after correcting for their radial velocity. The final spectra have a resolution of R $\sim$ 110\,000 and values of S/N that vary from 
as low as $\sim 20$ to as high as $\sim 820$, depending on the amount and quality of the original spectra. 

Figure \ref{fig1} shows the S/N distribution for the stars for which we were able to derive precise parameters. Although there is a significant amount of spectra have a low S/N, the vast 
majority ($\sim$ 75\%) have a S/N greater than 100 and $\sim$ 25\% a very good S/N higher than 200.

Our method for deriving spectroscopic stellar parameters is based on a differential analysis with respect to the Sun (using an adopted value of $log_{\epsilon}(Fe) = 7.47$)\footnote{This value taken from \citet[][]{Gonzalez-2000} was 
first used in \citet[][]{Santos-2000b}
and is adopted here to guarantee the homogeneity of our metallicities}, therefore there is some limitation on the spectral type of the stars for which we can derive good and precise stellar parameters. 
Our spectroscopic method is effective for F, G, and K stars with temperatures ranging from $\sim$ 4500 K to $\sim$ 6400 K. For this reason and since the sample is composed of a wider variety of spectral types, we made a 
pre-selection of the stars for which it would be appropriate to use our method to derive precise spectroscopic parameters. The selection was made by considering the index color B-V taken from the Hipparcos 
catalog \citep[][]{Hipparcus-2007}. We only selected the stars with B-V $<$ 1.2 to avoid the coolest stars in the sample. This B-V cutoff reduced the number of stars to 651.
Amoung these 651 stars there are a significant number (69) for which we were unable to derive good and precise parameters. The main reason for this is that the spectra of most of these stars have a very low S/N. There were also some 
cases of stars with high rotation that affected the spectra in a way that prevented an accurate derivation of spectroscopic parameters using our automatic methods (e.g. problems with strong blended lines). Among these 69 stars, 
there were also a few M stars that have passed the B-V cut-off and for which we cannot derive parameters. Altogether, the final number of stars with precise and homegeneously determined parameters is 582.

\section{Stellar parameters}

The spectroscopic stellar parameters and metallicities were derived following the same procedure used in previous works \citep[][]{Santos-2004b, Sousa-2006, Sousa-2008, Sousa-2011}. 
The method is based on the equivalent widths (hereafter EWs) of Fe I and Fe II weak lines, 
by imposing excitation and ionization equilibrium assuming LTE. We used the 2002 version of 
the code MOOG \citep[][]{Sneden-1973} and a grid of Kurucz Atlas 9 plane-parallel model atmospheres \citep[][]{Kurucz-1993}. In this procedure 
[Fe/H] is used as a proxy for the metallicity.
\begin{figure}[!t]
\centering
\includegraphics[width=8cm]{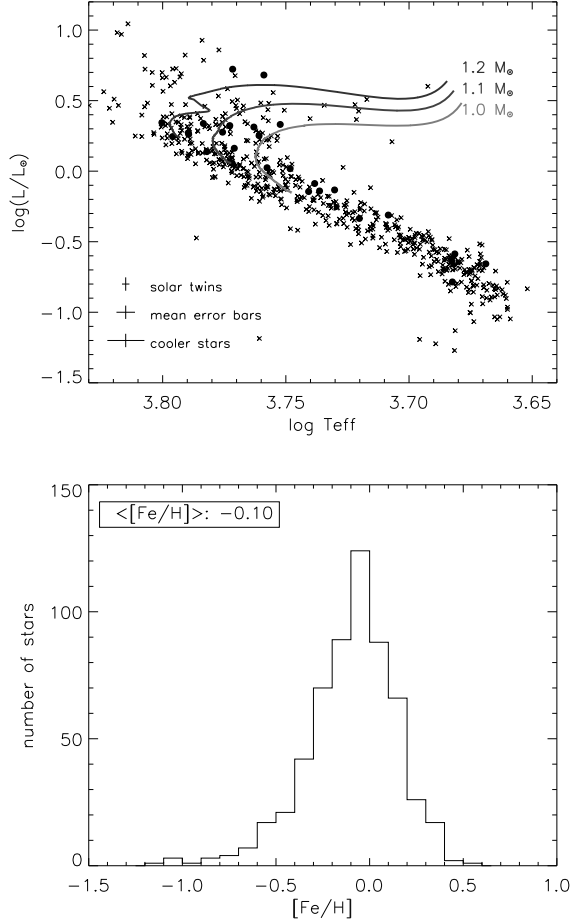}
\caption[]{In the top panel, we present the distribution of the sample stars in the H-R diagram. We also plot some evolutionary tracks computed with CESAM for a 1.0, 1.1, and 
1.2 M$_{\sun}$ star of solar metallicity. The filled circles represent the planet hosts in the sample. 
In the bottom panel, we present the metallicity distribution of this sample.}
\label{fighr}
\end{figure}
The EWs were automatically measured with the ARES\footnote{The ARES code can be downloaded at 
http://www.astro.up.pt/$\sim$sousasag/ares} code \citep[Automatic Routine for line Equivalent widths in stellar 
Spectra - ][]{Sousa-2007} that reproduces with success the common ``manual`` and interactive determination of EW measurements. The procedure used in \citet[][]{Sousa-2008} was closely followed and the same 
input parameters for ARES were used in this work. There is a significant number of stars in this sample with S/N lower than 100 and even a few spectra with S/N 
lower than 50, to which we applied the same procedure as discussed in \citet[][]{Sousa-2011}.

There are 14 stars in common between this volume-limited sample and the metal-poor sample analyzed in \citet[][]{Sousa-2011}. When comparing the results of these 14 stars, we obtained mean differences of 
$-0.5 \pm 0.6$ ($\sigma$ = 8.6) K for effective temperature; $0.01 \pm 0.00$ ($\sigma$ = 0.02) dex for surface gravity; and $0.00 \pm 0.00$ ($\sigma$ = 0.004) dex for [Fe/H]. This small 
mean differences demonstrates the high precision of our spectroscopic automatic analysis when using the same instrumental configurations.

The errors were determined in the same way as in previous works. We recall the discussion about precision versus accuracy errors presented in \citet[][]{Sousa-2011}. We present here precision errors, although 
in the electronic table we also add the ''accuracy`` error.

\subsection{Masses and luminosities}

Stellar masses were estimated following the same procedure as in \citet[][]{Sousa-2011}, where we applied the stellar evolutionary models from the Padova group 
using the web interface dealing with stellar isochrones and their derivatives \footnote{http://stev.oapd.inaf.it/cgi-bin/param}. We used 
the Hipparcos parallaxes and V magnitudes \citep[][]{Hipparcus-2007}, a bolometric correction from \citet[][]{Flower-1996}, and the effective temperature derived from the spectroscopic analysis. The errors 
presented for the masses were also acquired from  the web interface.

The luminosity was computed by considering the estimated Hipparcos parallaxes, V magnitude, and the bolometric correction. Its error is derived based on the parallax errors, which are the main source 
of uncertainty in the calculation of luminosity. The typical error in the luminosity is $\sim 0.05$, which was obtained assuming the mean parallax for the stars ($\sim$21 mas) and a typical error 
in the estimated parallaxes of 1.1 mas.

Figure \ref{fighr} presents some characteristics of the sample. The top plot shows the distribution of the sample on the Hertzsprung-Russell diagram, where we represent evolutionary tracks for
1.0, 1.1, and 1.2 M$_{\sun}$ stars of solar metallicity, computed using the CESAM code \citep[][]{Morel-1997, Marques-2008} \footnote{http://www.astro.up.pt/corot/models/}. Moreover, we present 
the typical error 
boxes on this specific diagram, where the error in the luminosity was described just before and the error in the temperature is derived from our spectroscopic method. 
This plot shows that the sample is composed mainly of main-sequence solar-type stars but also contains some subgiants. Two stars can be distinguished well below the main sequence. These stars are identified 
as HD 84627 and HIP 99606, being the second the farthest from the main sequence. 
We believe that the reason for these strange positions in the HR diagram comes directly from the parallax values that have indeed atypically large error values, $\sim 40 \pm 7$ $mas$
and $\sim 25 \pm 8$ $mas$, respectively, for HD 84627 and HIP 99606. The Hipparcos variability flag of HIP99606 is defined to be 3, meaning that variability is higher than 0.6 mag 
(the only one in the sample with flag set = 3). We indicate that HD84627 was found to have duplicity-induced variability in the Hipparcos catalog.

In the bottom panel of Figure \ref{fighr}, we present the metallicity distribution, which has a mean value of about $-0.10$ dex. As expected, this value is compatible with both 
comparison samples presented before in previous 
works \citep[][]{Santos-2004b,Sousa-2008} and the original sample of CORALIE, as can be seen in the work of \citet[][]{Santos-2004b} and later in this work in the top panel of Fig. \ref{fig_feh_planets}.

In Table \ref{tab2}, we present a subsample of the derived spectroscopic parameters. The full table will be available in electronic format.

\section{Internal comparisons}

To check the consistency of our determined parameters, we derived the effective temperature using two other independent methods developed in our previous works.

\begin{figure}[!t]
\centering
\includegraphics[width=8cm]{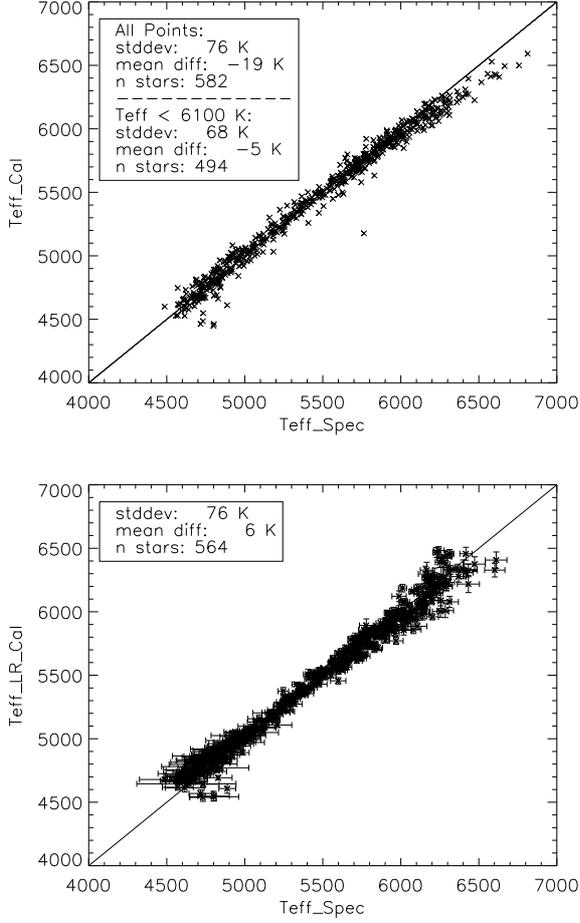}
\caption[]{Top panel: Comparison between the spectroscopically derived effective temperature and the temperature derived with a calibration based on B-V and [Fe/H]. Bottom panel: Comparison 
between the spectroscopically derived effective temperature and the temperature derived using line ratio calibrations obtained with the ``Line Ratio Calibration Code''.}
\label{figintcomp}
\end{figure}
\begin{figure}[!t]
\centering
\includegraphics[width=8cm]{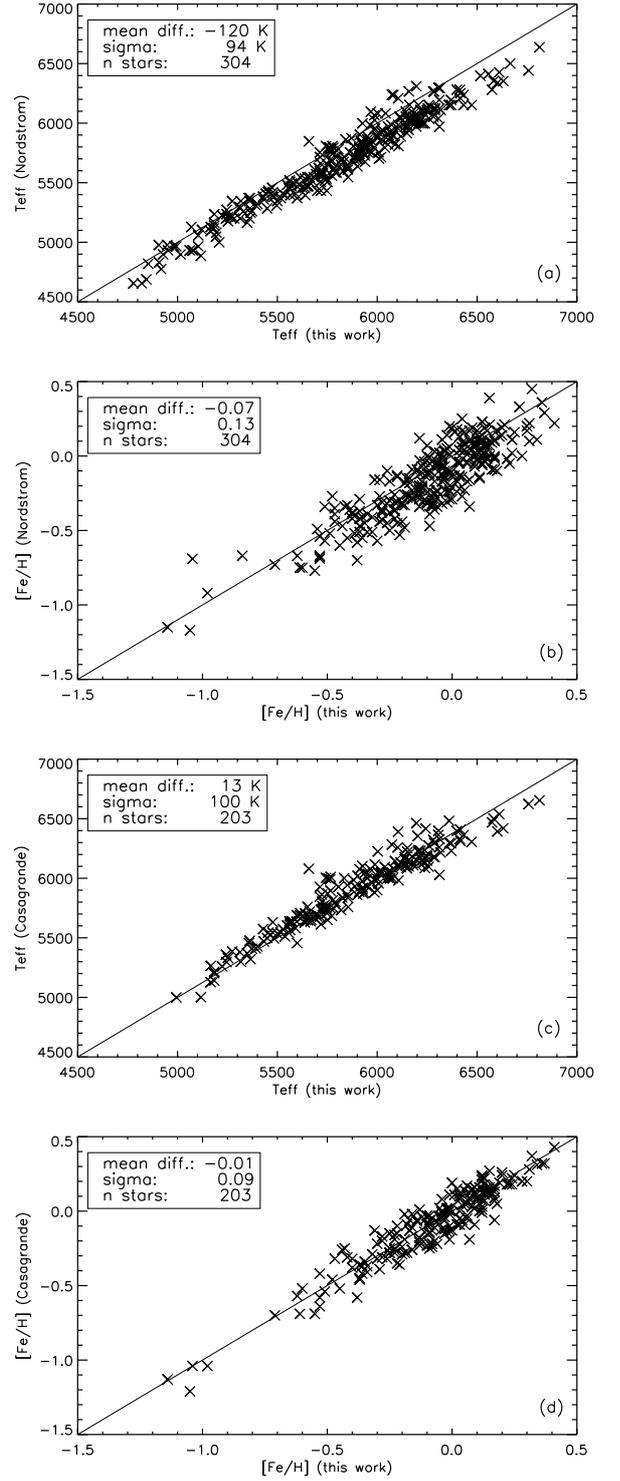}
\caption[]{Comparison between the spectroscopically derived effective temperature (a and c) and [Fe/H] (b and d) and those presented in \citet[][]{Nordstrom-2004} and \citet[][]{Casagrande-2011}, respectively.}
\label{fignord}
\end{figure}
\begin{figure}[!t]
\centering
\includegraphics[width=8cm]{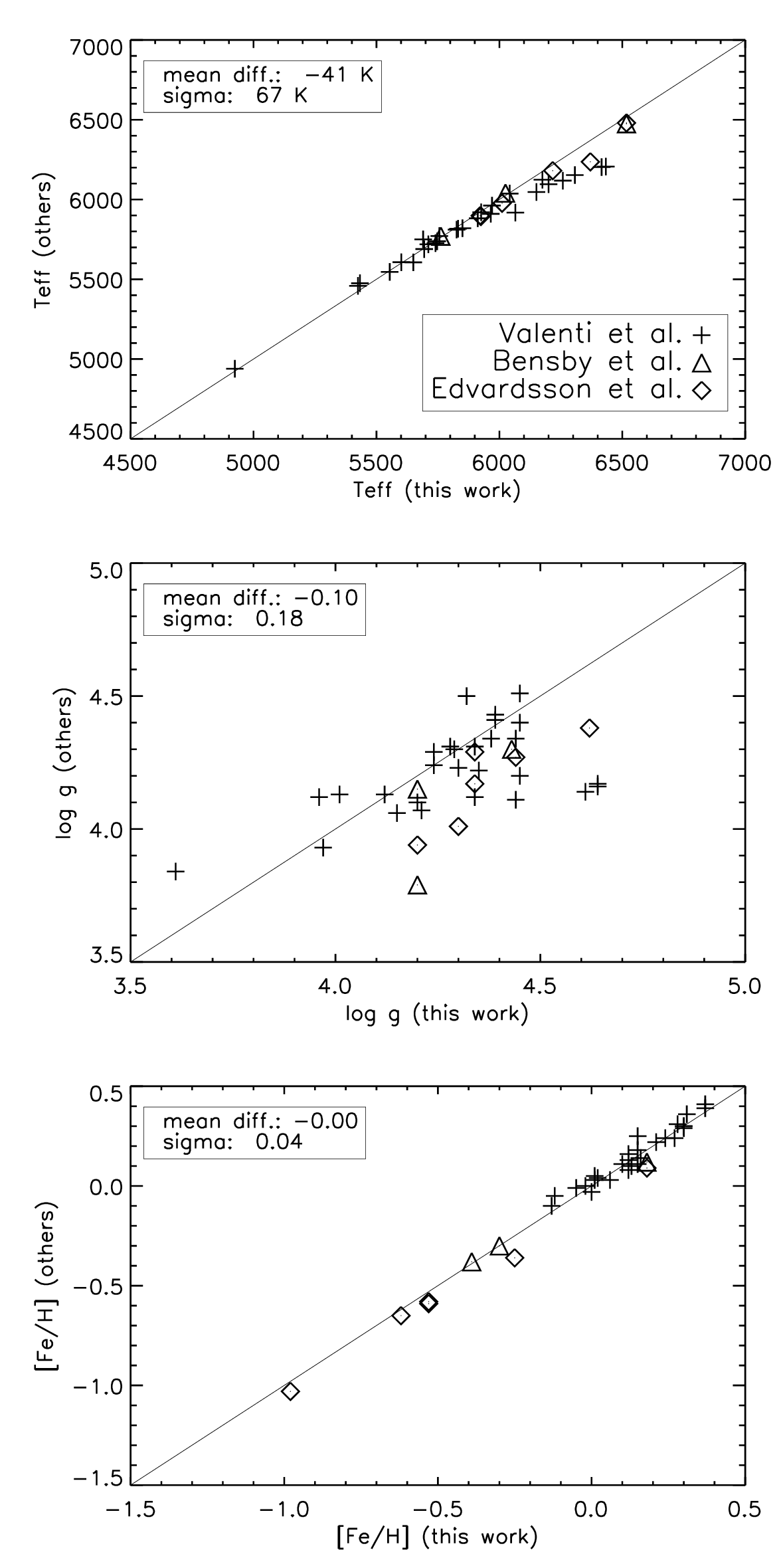}
\caption[]{Comparison between our determined spectroscopic stellar parameters, effective temperature (top panel), surface gravity (middle panel), and [Fe/H] (bottom panel) and the measurements of other authors. 
The crosses, triangles, and diamonds represent the comparison points with the values determined by \citet[][]{Valenti-2005}, \citet[][]{Bensby-2003}, and \citet[][]{Edvardsson-1993}, respectively.}
\label{figothers}
\end{figure}
We first used the latest calibration for the effective temperature as a function of \textit{B-V} and [Fe/H] derived in \citet[][]{Sousa-2011}. We used the color B-V from the Hipparcos 
catalogue and the spectroscopically derived [Fe/H]. The comparison between our spectroscopically determined effective temperature and the temperature derived using this calibration 
can be seen in the upper plot of Fig. \ref{figintcomp}. The temperatures are quite consistent between them having a mean difference (Teff\_Cal - Teff\_Spec) of only $-19$ K with a standard 
deviation of $76$ K. There is a 
slight offset for the hotter stars, but this offset can easily be explained by the calibration that we use only being valid for effective temperatures ranging from $4500 K$ to $6400 K$. This 
offset starts to be seen for stars with temperatures higher than 6100 K. When considering the stars with temperatures lower than 6100 K, we obtain a more relieble result with a mean difference of $-5$ K and a 
standard deviation of $68$ K. In this comparison there is a clear outlier, which is identified as HIP99606 an outlier also in the HR diagram. There might be a problem with the color B-V obtained from the Hipparcos catalogue for this star.

The second independent method that we used to check the consistency of our determined spectroscopic parameters was the one based on line ratios. We used ARES to derive the EW of the 
lines used to derive the line ratios. The estimation of the effective temperature was carried out using the freely available code ``Line Ratio Calibration Code'' accessible from the 
ARES webpage\footnote{The ''Line Ratio Calibration Code`` can be downloaded at http://www.astro.up.pt/$\sim$sousasag/ares}. The method behind this code and the respective determination of the calibrations for 
each line ratio is described in detail in \citet[][]{Sousa-2010}. There were some stars for which we could not derive the temperature using this method, for reasons similar to the one described before. There is also a 
range of temperatures that defines the validity of each line-ratio calibration. The typical range of temperatures is $[4500, 6400]$ K. Nevertheless, we were able to derive the effective temperature 
for 564 out of 582 stars with an impressive mean difference of only $6$ K and a standard deviation of $76$ K. This not only shows that our measured stellar parameters are consistent and on the same temperature 
scale as our previous measurements but again proves that the line-ratio code is very efficient and precise because it derived consistent temperatures for a sample of FGK stars that 
were not used for the calibration of the line ratios.

\section{Other studies}

We compared our results with those of other studies that have stars in common. This is useful to access the consistency between the various methods that use either spectroscopy or photometry to derive 
stellar parameters. We used the spectroscopic results of \citet[][]{Edvardsson-1993}, \citet[][]{Bensby-2003}, and \citet[][]{Valenti-2005}, and the photometric results of \citet[][]{Nordstrom-2004} and \citet[][]{Casagrande-2011}.

Figure \ref{fignord} compares our measurements of the effective temperature (panel (a)) and [Fe/H] (panel (b)) with those of \citet[][]{Nordstrom-2004}, which used 
the calibrations of \citet[][]{Alonso-1996} to derive the effective temperatures. In these plots, we have $\sim$ 300 stars in common. We can see a systematic offset between these two sets of results 
with a mean difference of $-120 \pm 5$ ($\sigma$ = 94) K. This offset is similar to that seen in \citet[][]{Sousa-2008}. 
The comparison between the derived [Fe/H] seems to show closer agreement, although there is still a mean difference of $-0.07 \pm 0.01$ ($\sigma$ = 0.13) dex that is significant.

In the same figure, we also compare our results with the photometry of \citet[][]{Casagrande-2011}. In this case, to make a robust comparison of the
two methods we only considered the stars in this work with more reliable photometry that are labeled with ''irfm'' pedigree (see \citet[][]{Casagrande-2011} for more details). In 
panel (c) of Figure \ref{fignord}, we can see the comparison 
between the two temperatures for more than 200 stars. The results shows a very low mean difference of only $13 \pm 7$ ($\sigma$ = 100) K and at higher temperatures a clear increase in the dispersion, 
which is expected as the stars are very different from the Sun and the errors increase accordingly. The metallicities presented in \citet[][]{Casagrande-2011}
are compared in panel (d) from the same figure. The comparison is quite consistent with a mean difference of only $-0.01 \pm 0.01$ ($\sigma$ = 0.09) dex.


\begin{table*}[t]
\caption[]{Solar twin candidates within the volume-limited sample observed with HARPS presented in this work.}
\begin{center}

\begin{tabular}{lccccccc}
\hline
\hline
Star ID     & T$_{\mathrm{eff}}$ & $\log{g}_{spec}$ & $\xi_{\mathrm{t}}$ & \multicolumn{1}{c}{[Fe/H]} & N(\ion{Fe}{i},\ion{Fe}{ii}) & $Age_{p}$  & $Mass_{p}$        \\
            & [K]                & [cm\,s$^{-2}$]   &  [km\,s$^{-1}$]    &                            &                             & [Gyr] & [M$_{\sun}$] \\
\hline

\object{HD\,19641  } &   5806	\ $\pm$\     61 & 4.39	\ $\pm$\   0.10 & 0.94\ $\pm$\  0.01 & -0.01\ $\pm$\  0.05 &  255, 35 &	4.6\ $\pm$\  3.4 & 0.992\ $\pm$\ 0.039  \\
\object{HD\,29263  } &   5780	\ $\pm$\     65 & 4.35	\ $\pm$\   0.10 & 0.94\ $\pm$\  0.03 &  0.03\ $\pm$\  0.05 &  259, 35 &	7.0\ $\pm$\  2.8 & 0.995\ $\pm$\ 0.030  \\
\object{HD\,76440  } &   5764	\ $\pm$\     63 & 4.43	\ $\pm$\   0.10 & 0.89\ $\pm$\  0.02 & -0.01\ $\pm$\  0.05 &  260, 35 &	3.7\ $\pm$\  3.2 & 0.983\ $\pm$\ 0.038  \\
\object{HD\,96116  } &   5832	\ $\pm$\     63 & 4.52	\ $\pm$\   0.10 & 0.96\ $\pm$\  0.03 & -0.01\ $\pm$\  0.05 &  260, 35 &	2.7\ $\pm$\  2.5 & 1.005\ $\pm$\ 0.036  \\
\object{HD\,134702 } &   5782	\ $\pm$\     64 & 4.50	\ $\pm$\   0.11 & 0.74\ $\pm$\  0.04 & -0.04\ $\pm$\  0.05 &  259, 36 &	3.2\ $\pm$\  3.0 & 0.979\ $\pm$\ 0.037  \\
\object{HD\,145927 } &   5819	\ $\pm$\     62 & 4.41	\ $\pm$\   0.10 & 0.93\ $\pm$\  0.02 & -0.03\ $\pm$\  0.05 &  254, 35 &	2.7\ $\pm$\  2.5 & 0.994\ $\pm$\ 0.036  \\
\object{HD\,216008 } &   5773	\ $\pm$\     62 & 4.38	\ $\pm$\   0.10 & 0.91\ $\pm$\  0.02 & -0.04\ $\pm$\  0.05 &  257, 35 &	6.1\ $\pm$\  3.9 & 0.967\ $\pm$\ 0.037  \\

\hline
\end{tabular}
 
\end{center}
\tablefoot{$\log{g}_{spec}$ the spectroscopic surface gravity; $\xi_{\mathrm{t}}$ is the microturbulance speed; N(\ion{Fe}{i},\ion{Fe}{ii}) is the number of lines used 
in the spectroscopic analysis; $Mass_{p}$ and $Age_{p}$ are the mass and age determined directly from the Padova web interface using the Hipparcos parallax;
The error presented here are the assumed ``accuracy'' errors, which were used to select solar twin candidates within the full sample.
}
\label{tab2}
\end{table*}

Figure \ref{figothers} shows a comparison between our different spectroscopically derived parameters, effective temperature (top panel), surface gravity (middle panel), and [Fe/H] (lower panel) and 
other spectroscopic values derived by other works, namely \citet[][]{Edvardsson-1993} (6 stars in common), \citet[][]{Bensby-2003} (3 stars in common), and \citet[][]{Valenti-2005} (28 stars in common).

For effective temperatures, we observe a mean difference of $-41 \pm 11$ ($\sigma$ = 67) K meaning that our temperatures are on average cooler. There seems to be a small offset for the hotter stars 
(with Teff $>$ 6100 K) derived by \citet[][]{Valenti-2005}. The reason for this offset is unclear. It is also present in a similar comparison made 
in \citet[][]{Sousa-2008} but is not visible when comparing with the results of other methods considered in the same work.

When comparing the surface gravity, we obtain a cloud of points around the identity line that is rather typical of surface gravities determined with different methods. The 
comparison finds a mean difference of $-0.10 \pm 0.03$ ($\sigma$ = 0.18) dex. The values derived by \citet[][]{Edvardsson-1993}, which used the Balmer discontinuity index, seems to be systematic lower than our 
determination as already pointed out in \citet[][]{Sousa-2008}.

Finally, the comparison of [Fe/H] that can be seen in the lower plot of Fig. \ref{figothers} shows a very good consistency with a mean difference of $0.00 \pm 0.01$ ($\sigma$ = 0.04) dex. 
This shows that the different authors have very consistent metallicities measurements despite the differences in surface gravity and temperature, for the hotter stars.

All these comparisons confirm the consistency of the parameters derived by our team.

\section{Solar twins in the sample}

The Sun is by far the most well-known star. But how similar are other stars to the Sun. To understand this question, a good starting point is to compare the Sun to similar stars. 
With this comparison, we can start to ask questions such as: How frequent are the stars in the solar neighborhood with the same mass, temperature, and metallicity?
How different is their chemical abundance? Do they also host planets? Are these planets similar to the ones we find in our Solar System?

The quest to find solar twins has been ongoing for a long time \citep[][]{Cayrel_de_Strobel-1996}. Here we present a few new solar-twin candidates within the volume-limited sample observed with HARPS.
The list of star candidates to be solar twins are listed in Table \ref{tab2}. They were selected such that their ``accuracy error'' (given in Sect. 3) include the solar values for 
each parameter (5777K for effective temperature, 4.44 dex for logarithmic surface gravity, 0.00 dex for [Fe/H], and 1.0 for mass). From the table, one can see that the age estimates are also within the solar age.

\section{The largest volume-limited sample}

\subsection{[Fe/H] based on the CCF calibration}

\begin{figure}[!t]
\centering
\includegraphics[width=8cm]{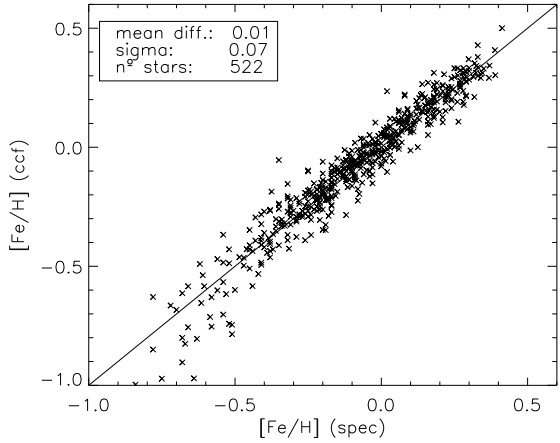}
\caption[]{Comparison between the [Fe/H] derived from the CORALIE CCF and the [Fe/H] derived using our spectroscopic method.}
\label{figccf}
\end{figure}

As already indicated in previous sections, the sample for which we present parameters is an extension of the CORALIE planet-search sample \citep[][]{Udry-2000}. This sample was presented as a comparison 
sample in \citet[][]{Santos-2004b}, where the metallicities of all the stars were determined using a cross correlation function (CCF) calibration. We have already a spectroscopic determination for a 
significant number of stars 
within the CORALIE sample for which we derived precise spectroscopic parameters using HARPS spectra analysed in previous work for other different planet 
search samples \citep[][]{Santos-2004b, Santos-2005, Sousa-2006, Sousa-2008, Sousa-2010}. In 
Fig. \ref{figccf}, we compare our [Fe/H] spectroscopic measurements with the one derived using the CCF calibration for stars in the CORALIE sample. The results shown in the figure for the 522 stars with a 
spectroscopically determined metallicity are very consistent and have a mean difference of only $0.01 \pm 0.00$ ($\sigma$ = 0.07) dex. This small mean difference is negligible (when 
considering the typical [Fe/H] error for each star) and 
its consistency allows us to use the CCF metallicities, for the stars without our spectroscopic derivation, to construct 
the largest uniform volume-limited sample by combining both the HARPS and CORALIE samples. This includes most FGK dwarfs in a volume of $\sim$ 60 pc around the Sun.

Although the consistency between our spectroscopic [Fe/H] and the one derived using the CCF calibration is very good, we use the more precise spectroscopic derivation for stars when one is available. 
This only concerns the CORALIE sample that we use here and is composed of a total of $\sim$ 1250 stars for which 522 stars have spectroscopically derived [Fe/H] in our previous works. When 
combining these stars with the HARPS sample presented in this work, we achieve a total number of stars of 1830 (1104 out of 1830 with homogeneous spectroscopic metallicity measurements). 
This ensures that it is the largest sample of stars with precise and homogeneous [Fe/H] measurements ever studied in the framework of planet formation and evolution.

\subsection{The frequency of planets with [Fe/H]}
\begin{figure*}[t]
\centering
\includegraphics[width=18cm]{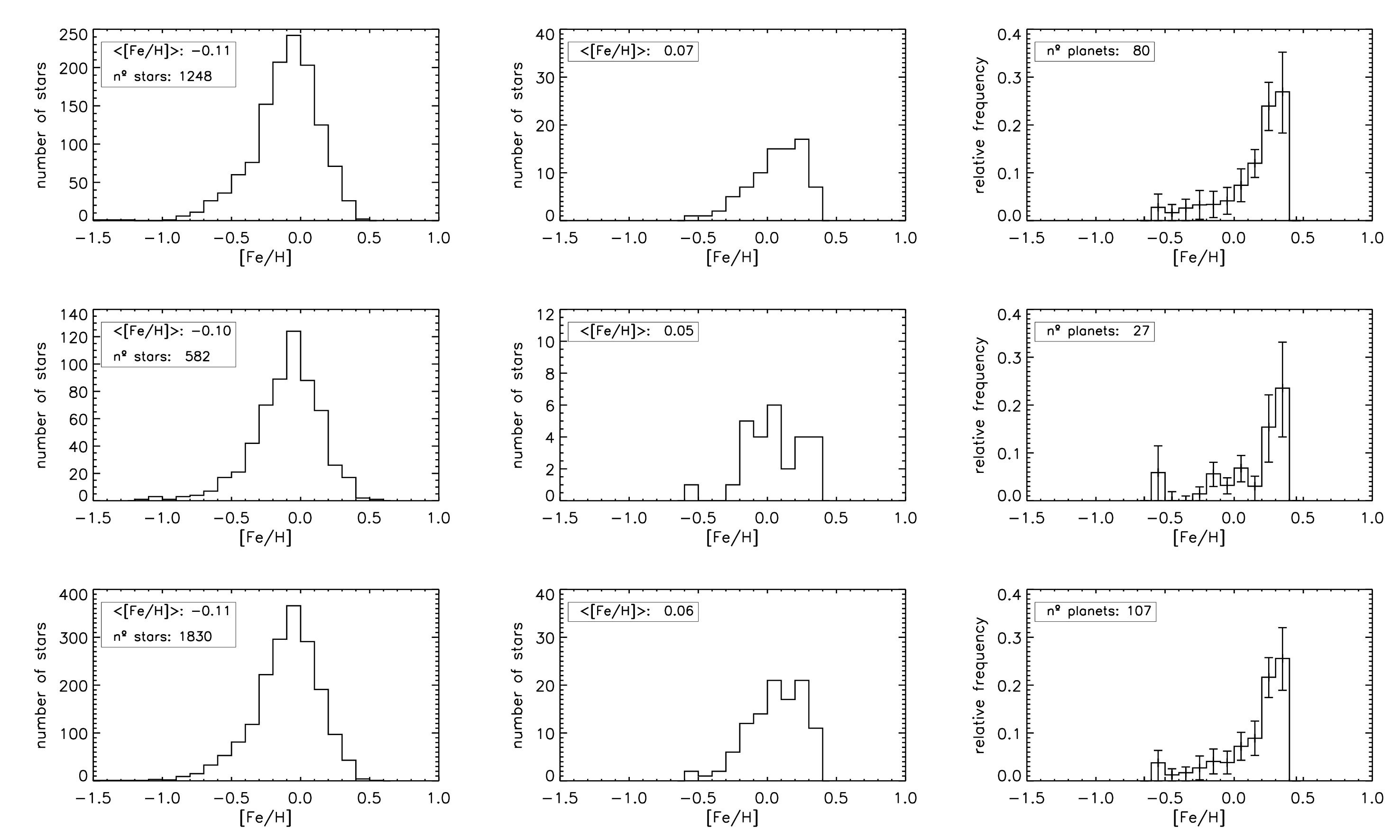}
\caption[]{Metallicity distributions for the CORALIE sample (top panels), the HARPS sample presented in this work (middle panels), and the union of the two samples (bottom panels). 
The left panels shows the full samples distribution of [Fe/H], while the center plots for each sample shows the planet host [Fe/H] distribution. The right plots shows the frequency of planets for each [Fe/H] bin.}
\label{fig_feh_planets}
\end{figure*}


In Fig. \ref{fig_feh_planets}, we present metallicity distributions for the initial CORALIE sample (top panels), the HARPS samples discussed in this work (middle panels), and the large and combined sample (lower panels). 
The middle and right plots correspond to the population of planet host stars in each sample. There are already a few planet hosts presented in the HARPS volume-limited sample that have been discovered in the last 
years \citep[][]{Tinney-2003, Hebrard-2010, Santos-2010b, Naef-2010, Moutou-2009, LoCurto-2010, Santos-2011, Mordasini-2011, Moutou-2011}. For all of the planet hosts, we were careful to include only the stars 
that have confirmed planets. In this way, we obtained a total of 107 planet hosts, 80 in the CORALIE sample and 27 in the HARPS sample. The information about the presence of planets was updated using the exoplanet 
encyclopedia\footnote{http://exoplanet.eu/} and confirmed by the references therein.

Looking at the top panels, for the CORALIE sample, the left panel shows the typical metallicity distribution for the solar neighborhood. 
This distribution is very similar to the ones presented in studies dealing specifically with the metallicity distribution in the solar neighborhood which are based on unbiased, volume-limited 
samples \citep[e.g.][]{Haywood-2001, Fuhrmann-2004, Fuhrmann-2008, Casagrande-2011}. The CORALIE distribution has a similar shape indicating that this sample is also rather 
unbiased and representative of the stellar population in the solar neighborhood. 

The distribution of the planet hosts is, as seen before, very different from that of the full 
sample, being more metal rich. The probability of belonging to the same distribution according to a Kolmogorov-Smirnov test (hereafter PKS) is nearly zero (PKS: $2.3 e^{-8}$). The right plot shows the frequency of planet host for each bin of 
metallicity. This plots shows an evident higher probability of finding giant planets around metal-rich stars.

The middle plots of Fig. \ref{fig_feh_planets} displays the same sequence of plots for the HARPS sample. This sample shows the same kind of the metallicity distribution as seen 
for the CORALIE sample. We find that 
both full samples have a very significant PKS of $\sim 83 \%$. On the other hand, although the HARPS planet-host metallicity distribution is very different from that of the full HARPS 
sample (with a PKS of $\sim 2\%$) it has a somewhat different shape from the distributions in the top plot for the the CORALIE planet hosts. Their PKS is $\sim 14\%$, a value that indicates that there 
are some differences between the two samples. However, when we use these data to plot the planet frequency for each metallicity bin within the sample (right panels of Fig. \ref{fig_feh_planets}) 
we can see that the plots are again quite similar, i.e. 
in both plots an increase in the planet frequency can clearly be seen for the metal-rich stars.

Finally, in the bottom panels we see the same sequence of plots for both samples combined. The difference in the mean metallicities between the planet hosts and the full sample is 0.15 dex. 
The same trends observed before 
for each individual sample are also observed here, showing again the general correlation with metallicity that implies the giant planets are more likely to be discovered around metal-rich stars.

The error bars expressed in the relative frequency plots were derived assuming a Gaussian distribution when the total number of stars in a bin was higher or equal to 100 or a Poisson distribution when lower than 100 stars.
\begin{figure}[t]
\centering
\includegraphics[width=8cm]{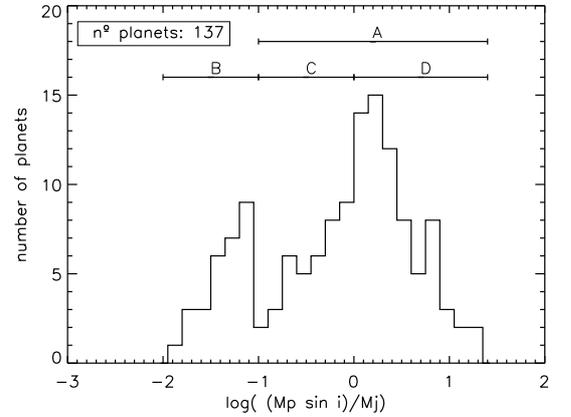}
\caption[]{Planetary mass distribution in log scale, illustrating the two evident peaks at two distinct planet mass regimes, the neptunian mass and the jupiter mass planets.}
\label{fig_massp}
\end{figure}
\subsection{Planet frequency, planetary mass, and [Fe/H]}
\begin{table*}[t]
\caption[]{Number and frequency of stars and planet hosts for [Fe/H] bin considering the mass of the most massive planets present.}
\begin{center}

\begin{tabular}{|l|r|rl|rl|rl|rl|rl|}
\hline
[Fe/H] bin & nstars & np & fnp (full) & np & fnp \textbf{(A)} & np & fnp \textbf{(B)} & np & fnp \textbf{(C)} & np & fnp \textbf{(D)}\\
\hline
\hline
$[-0.6, -0.5[$ & 53 & 2 &( 3.77 \%) & 2 &( 3.77 \%) & 0 &( 0.00 \%) & 2 &( 3.77 \%) & 0 &( 0.00 \%)\\
$[-0.5, -0.4[$ & 81 & 1 &( 1.23 \%) & 1 &( 1.23 \%) & 0 &( 0.00 \%) & 0 &( 0.00 \%) & 1 &( 1.23 \%)\\
$[-0.4, -0.3[$ &118 & 2 &( 1.69 \%) & 0 &( 0.00 \%) & 2 &( 1.69 \%) & 0 &( 0.00 \%) & 0 &( 0.00 \%)\\
$[-0.3, -0.2[$ &222 & 6 &( 2.70 \%) & 4 &( 1.80 \%) & 2 &( 0.90 \%) & 2 &( 0.90 \%) & 2 &( 0.90 \%)\\
$[-0.2, -0.1[$ &296 &12 &( 4.05 \%) &11 &( 3.72 \%) & 1 &( 0.34 \%) & 5 &( 1.69 \%) & 6 &( 2.03 \%)\\
$[-0.1, +0.0[$ &336 &14 &( 4.17 \%) &11 &( 3.27 \%) & 3 &( 0.89 \%) & 4 &( 1.19 \%) & 7 &( 2.08 \%)\\
$[+0.0, +0.1[$ &291 &21 &( 7.22 \%) &21 &( 7.22 \%) & 0 &( 0.00 \%) &10 &( 3.44 \%) &11 &( 3.78 \%)\\
$[+0.1, +0.2[$ &191 &17 &( 8.90 \%) &16 &( 8.38 \%) & 1 &( 0.52 \%) & 4 &( 2.09 \%) &12 &( 6.28 \%)\\
$[+0.2, +0.3[$ & 97 &21 &(21.65 \%) &20 &(20.62 \%) & 1 &( 1.03 \%) & 6 &( 6.19 \%) &14 &(14.43 \%)\\
$[+0.3, +0.4[$ & 43 &11 &(25.58 \%) &11 &(25.58 \%) & 0 &( 0.00 \%) & 4 &( 9.30 \%) & 7 &(16.28 \%)\\
\hline
Total planets: & & 107 &  & 97 & & 10 & & 37 & & 60 & \\
$<Fe/H>$: & -0.11 & & 0.06 & & 0.08 & & -0.11&  & 0.04 &  & 0.10 \\
\hline
\end{tabular}
\end{center}
\label{tab3}
\end{table*}

\begin{figure*}[t]
\centering
\includegraphics[width=18cm]{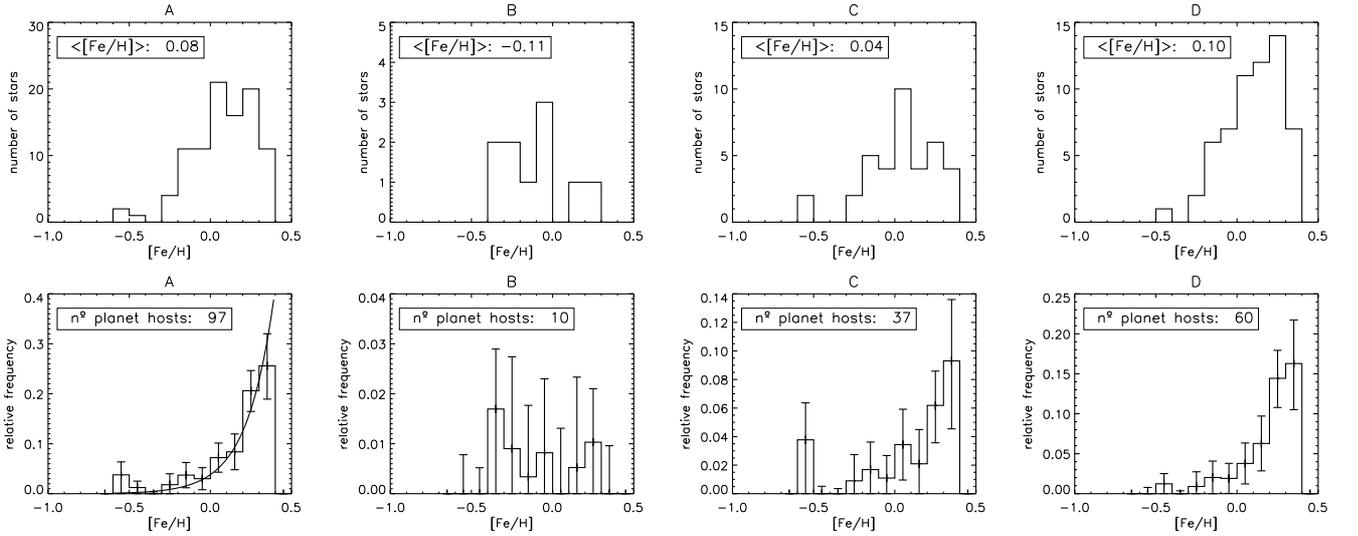}
\caption[]{Metallicity distribution and respective frequency for the planet host with the four different planet-mass regimes described in Figure \ref{fig_massp}. In the left 
and bottom plot, we fit a power law to the data - see more details in the text.}
\label{fig_massp_freq}
\end{figure*}
We analyze the planet frequency around stars considering the mass of the present planets. In Figure \ref{fig_massp}, we show the distribution of the mass of the planets present around the 
stars in the large HARPS + CORALIE sample. The presence of two visible peaks in different mass regimes is clearly visible, namely the Jovian planets and the Neptune-mass planets. This was previously found in the 
review of \citet[][]{Udry-2007}. In the same figure, we show the intervals of mass of planets that we analyze. We consider four intervals of masses. The Jupiter-mass planets with masses between 0.1 and 25 
Jupiter masses (mass interval A). The Neptune-mass planets with masses between 0.01 and 0.1 Jupiter masses (mass interval B). The Jovian planets are also sub-divided into the lighter jovians with masses between 0.1 and 1 
Jupiter masses (mass interval C) and heavy jovians with masses between 1 and 25 Jupiter masses (mass interval D).

In Figure \ref{fig_massp_freq}, we present the metallicity distribution and the planet frequency for each of the mass intervals described. The frequencies of planets in each interval are also presented in Table \ref{tab3}. For this 
analysis and for the stars hosting planetary systems, we only consider the mass of the most massive planet present in the system.

For interval A, which contains stars with giant planets we can again see the higher frequencies for the more metal.rich stars. In this case, we fit a power law to the histogram of the planet frequency. The 
fitted function is giving by:
\[
P_{(planet)} = 0.038 [(N_{Fe}/N_H)/(N_{Fe}/N_H)] ^{2.58}, 
\]
which is not very different from the one presented in \citet[][]{Valenti-2005} with the coefficients (0.03, 2). Although in our case, the fitted power law does not appear to reproduce the very rapid 
increase in frequency 
around the 0.2 [Fe/H] bin. This rapid increase is clearly evident, and may represent a discontinuity in the planet formation efficiency as a function of [Fe/H].
In the same figure, we can see the metallicity distribution and planet frequency for the other intervals (B, C, and D). It is interesting to see a gradual increase in the mean metallicity 
of each interval as we increase the mass of the planets.
This may indicate that there is some superposition of the two groups of planets in the mass bins.
We can also infer from this observational result that the metallicity distribution of the Neptunian hosts is rather flat compared to that of the stars hosting jovians. This evidence illustrates once again the 
possible difference between the formation of Neptune-mass planets and Jupiter-mass planets as already noted in \citet[][]{Sousa-2008},and \citet[][]{Udry-2007}, which is supported by theoretical models based
 on the core accretion idea \citep[][]{Mordasini-2009}.

These results should be intrepreted with caution because observational bias may be present at some level. On the one hand, the most successful methods such as the use of radial velocities or transits 
are biased in their detections because it is easier to find more massive planets and those closer to the host star. On the other hand, several planet-search programs are focused on specific samples, 
constrained to contain only high metal-content stars, or using observational strategies designed to identify short period planet. All these together can introduce observational biases into the planet host subsamples.



\section{Summary}

We have presented precise and homogeneous measurements of spectroscopic stellar parameters for a volume-limited sample of stars observed within the HARPS planet-search program.
The parameters were derived with a method similar to one we previously used to derive a set of homogeneous and precise parameters for all the stars. We estimated the mass of each star 
using the Padova web interface.

We have verified the spectroscopically derived temperature using two other independent calibrations, one using the B-V color and measurement of [Fe/H] and the other using calibrated line-ratios. 
Both calibrations infer very 
consistent temperatures. We have compared our derived parameters with others found in the literature. The comparisons performed here show compatible results, with some small offsets that were briefly discussed.

We have combined the HARPS sample presented here with the original CORALIE sample. The latter sample is composed of stars that have either a spectroscopically determined [Fe/H] derived in our previous works, 
or a CCF-calibrated [Fe/H] that is proven to be compatible with our spectroscopic [Fe/H]. Both samples together constitute the largest volume-limited sample with homogeneous and precise [Fe/H].

Using the largest volume-limited metallicities, we have performed a simple analysis of the planet host distributions, focusing also on the masses of the planets present in each system. In this 
analysis we have confirmed again the higher frequency of planets for metal-richer stars, but we have also found a more rapid increase in frequency around 0.2 dex.

After dividing the system accordingly to the mass of the most massive planet orbiting the planet host, we have found a clear trend with metallicity. Stars hosting less massive planets have on average 
lower metallicity than stars 
with higher planet mass. This is clearly seen when dividing the system into three bins of masses: Neptune regime, lighter Jovians regime, and heavy Jovian regime. This result also should be interpreted 
with care because of possible observational bias present in the subsample containing the planet host.

\begin{acknowledgements}
S.G.S acknowledges the support from the Funda\c{c}\~ao para a Ci\^encia e Tecnologia (Portugal) in the form of a grants SFRH/BPD/47611/2008. NCS thanks for the 
support by the European Research Council/European Community under the FP7 through a Starting Grant, as well as the support from Funda\c{c}\~ao para a Ci\^encia 
e a Tecnologia (FCT), Portugal, through programme Ci\^encia\,2007. We also acknowledge support from FCT in the form of grants reference PTDC/CTE-AST/098528/2008, 
PTDC/CTE-AST/66181/2006, and PTDC/CTE-AST/098604/2008.

\end{acknowledgements}

\bibliographystyle{/home/sousasag/posdoc/mypapers/aa-package/bibtex/aa}
\bibliography{sousa_bibliography}

\end{document}